\documentclass[conference]{IEEEtran}

\usepackage{cite}
\usepackage{amsmath,amssymb,amsfonts}
\usepackage{algorithm,algorithmic}
\usepackage{graphicx}
\usepackage{textcomp}
\usepackage{xcolor}
\usepackage{subfigure}
\usepackage{fancyhdr}
\fancypagestyle{firstpage}{
\fancyhf{}
\fancyfoot[L]{978-1-7281-5409-1/20/\$31.00 \copyright 2020 IEEE.}

}
\def\BibTeX{{\rm B\kern-.05em{\sc i\kern-.025em b}\kern-.08em
    T\kern-.1667em\lower.7ex\hbox{E}\kern-.125emX}}

\renewcommand{\baselinestretch}{0.975}
    
\begin{document}

\title{SANSCrypt: A Sporadic-Authentication-Based Sequential Logic Encryption Scheme\\
\thanks{Identify applicable funding agency here. If none, delete this.}
}

\author{
Yinghua Hu, Kaixin Yang, Shahin Nazarian, and Pierluigi Nuzzo\\ 
Department of Electrical and Computer Engineering \\
University of Southern California, Los Angeles, CA, USA \\ \{yinghuah, kaixinya, shahin.nazarian, nuzzo\}@usc.edu\\[0ex]
}
{\onecolumn \textcopyright 2020 IEEE. Personal use of this material is permitted. Permission from IEEE must be obtained for all other uses, in any current or future media, including reprinting/republishing this material for advertising or promotional purposes, creating new collective works, for resale or redistribution to servers or lists, or reuse of any copyrighted component of this work in other works. }
\twocolumn
\maketitle
\thispagestyle{firstpage}
\begin{abstract}
We propose SANSCrypt, a novel sequential logic encryption scheme to protect integrated circuits against reverse engineering. Previous sequential encryption methods focus on modifying the circuit state machine such that the correct functionality can be accessed 
by applying the correct key sequence only once. Considering the risk associated with one-time authentication, SANSCrypt adopts a new temporal dimension to logic encryption, by requiring the user to sporadically perform multiple authentications according to a protocol based on pseudo-random number generation. Analysis and validation results on a set of benchmark circuits show that SANSCrypt offers a substantial output corruptibility if the key sequences are applied incorrectly. Moreover, it exhibits an exponential resilience to existing attacks, including SAT-based attacks, while maintaining a reasonably low overhead. 
\end{abstract}


\section{Introduction}
\label{sec:intro}

The design process of modern VLSI systems
often relies on a supply chain where several services, such as verification, fabrication, and testing, are outsourced to third-party companies. If these companies gain access to a sufficient amount of critical design information, they can potentially reverse engineer the design. 
One possible consequence of reverse engineering is Hardware Trojan (HT) insertion, which can be destructive for many applications. HTs
can either disrupt the normal circuit operation~\cite{karri2010trustworthy} or provide the attacker with access to critical data or software running on the chip~\cite{tehranipoor2010survey}.

Countermeasures such as logic encryption~\cite{rajendran2013fault, yasin2017provably, yasin2016sarlock, chakraborty2009harpoon}, integrated circuit (IC) camouflaging~\cite{yasin2016camoperturb}, watermarking~\cite{charbon1998hierarchical}, and split manufacturing~\cite{xiao2015efficient} have been developed over the past decades to prevent IC reverse engineering. 
Among these, logic encryption has received significant attention as a promising, low-overhead countermeasure. Logic encryption modifies the circuit in a way such that
a user can only access the correct circuit functionality after providing a correct key sequence. 
Otherwise, the circuit function remains hidden, and the output  different from the correct one.

Various logic encryption techniques~\cite{rajendran2013fault, yasin2017provably, yasin2016sarlock, chakraborty2009harpoon} and potential attacks~\cite{subramanyan2015evaluating, chakraborty2019surf, shen2019sigattack} have appeared in the literature, as well as methods to systematically evaluate them~\cite{vivek2019system, hu2019models}. A category of techniques~\cite{rajendran2013fault, yasin2017provably, yasin2016sarlock} is designed to modify and protect the combinational logic portions of the chip and can be extended to sequential circuits by assuming that the scan chains are not accessible by the attacker, e.g., due to scan chain encryption and obfuscation~\cite{sengar2007secured, paul2007vim, wang2017secure}. 
Another category of techniques, namely, sequential logic encryption~\cite{chakraborty2009harpoon,desai2013interlocking,kasarabada2019deep}, targets, instead, the state transitions of the original finite state machine (FSM). Sequential logic encryption introduces additional states and transitions in the original FSM, essentially partitioning the state space into two sets. After being powered on or reset, the FSM enters the \textit{encrypted mode}, exhibiting an incorrect output behavior. The FSM transitions, instead, to the \textit{functional mode}, providing the correct functionality, upon receiving a sequence of key patterns.

A set of attacks have been reported against sequential encryption schemes, aiming to retrieve the correct key sequence or circuit function. Shamsi~\textit{et al}.~\cite{shamsi2019kc2} adapted the Boolean satisfiability (SAT)-based attack~\cite{subramanyan2015evaluating}, traditionally targeted to combinational logic encryption, by leveraging methods from bounded model checking to unroll the sequential circuit. 
Recently, an attack based on automatic test pattern generation (ATPG)~\cite{duvalsaint2019atpg} uses concepts from excitation and propagation of stuck-at faults to search the key sequence 
among the input vectors generated by ATPG. 
When the attackers have some knowledge of the topology of the encrypted FSM, then they can extract and analyze  the state transition graph and bypass the encrypted mode~\cite{meade2017revisit}. 
Overall, the continuous advances in FSM extraction and analysis tools tend to challenge any of the existing sequential encryption
schemes and call for approaches that can significantly increase their robustness. 

This paper proposes \emph{SANSCrypt}, a Sporadic-Authentication-based Sequential Logic Encryption (SANSCrypt) scheme, which raises the attack difficulty via a \emph{multiple-authentication protocol}, whose decryption relies on \emph{retrieving a set of key sequences as well as the time at which the sequences should be applied}. 
Our contributions can be summarized as follows: 
\begin{itemize}
	\item A robust, multi-authentication based sequential logic encryption
method that for the first time, to the best of our knowledge, systematically incorporates
the robustness of multi-factor authentication (MFA)~\cite{bhargav2007privacy} in the context of hardware obfuscation. 
    \item An architecture for sporadic re-authentication where key sequences must be applied at multiple random times, determined by a random number generator, to access the correct circuit functionality. 
    \item Security analysis and empirical validation of SANSCrypt on a set of ISCAS'89 benchmark circuits~\cite{brglez1989combinational}, showing exponential resilience against existing attacks, including sequential SAT-based attacks, and reasonably low overhead. 
\end{itemize}

\noindent Analysis and validation results show that SANSCrypt can significantly enhance the resilience of sequential logic encryption under different attack assumptions. 

\section{Background and Related Work}\label{sec:background}

Among the existing sequential logic encryption techniques, HARPOON~\cite{chakraborty2009harpoon} 
defines two modes of operation.  
When powered on, the circuit is in the encrypted mode and exhibits an incorrect functionality.  
The user must apply a sequence of input patterns 
during the first few clock cycles 
to enter the functional mode, in which the correct functionality is recovered. 
However, 
the encrypted mode and functional mode FSMs are connected by only one transition (edge), which can be exploited by an attacker to perform FSM extraction and analysis,
and  
bypass the encrypted mode~\cite{meade2017revisit}. 

Interlocking~\cite{desai2013interlocking} sequential encryption modifies the circuit FSM such that multiple paths are available between the states of the encrypted and the ones of the functional FSMs, making it harder for the attacker to detect the only correct transition between the two modes.  
However, in both HARPOON and Interlocking encryption, once the circuit enters the functional mode, it remains there until reset. 

Dynamic State-Deflection~\cite{dofe2018novel} requires, instead, an additional key input verification step while in the functional mode. If the additional key input is incorrect, the FSM transitions to a black-hole state cluster which can no longer be left. However, because the additional key input is fixed over time, the scheme becomes more vulnerable to sequential SAT-based attacks~\cite{shamsi2019kc2}.

Finally, instead of corrupting the circuit function immediately after reset,  DESENC~\cite{kasarabada2019deep} counts the occurrence of a specific but rare event in the circuit. Once the counter reaches a threshold, the circuit enters the encryption mode. This scheme is more resilient to sequential SAT-based attacks~\cite{kasarabada2019sequentialsat} because it requires unrolling the circuit FSM a large number of times to find the key. However, the initial transparency window may still expose critical portions of the circuit functionality.

\section{SANSCrypt}\label{sec:theory}

We introduce design and implementation details for SANSCrypt, starting with the underlying threat model.

\subsection{Threat Model}

SANSCrypt assumes a threat model that is consistent with the previous literature on  sequential logic encryption~\cite{chakraborty2009harpoon, shamsi2019kc2, meade2017revisit}. The goal of the attack is to access the correct circuit functionality, by either reconstructing the deobfuscated circuit or finding the correct key sequence. To achieve this goal, the attacker can leverage one or more of the following resources: (i) the encrypted netlist;  
(ii) a working circuit providing correct input-output pairs; (iii) knowledge of the encryption technique. In addition, we assume that the attacker has no access to the scan chain and cannot directly observe or change the state of the circuit. 

\subsection{Authentication Protocol}

\begin{figure}[t]
\centerline{
\subfigure[]{
\includegraphics[width=0.45\columnwidth]{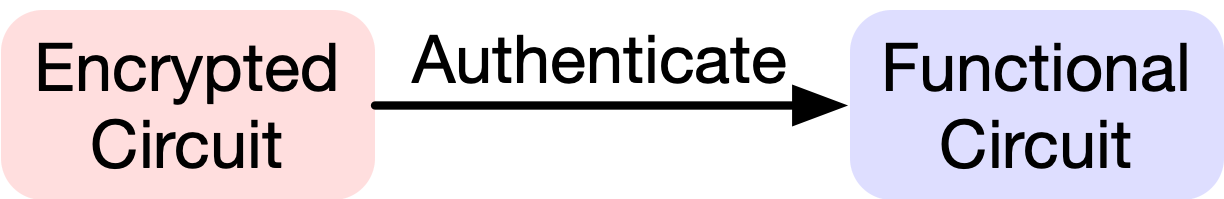}
}
\subfigure[]{
\includegraphics[width=0.45\columnwidth]{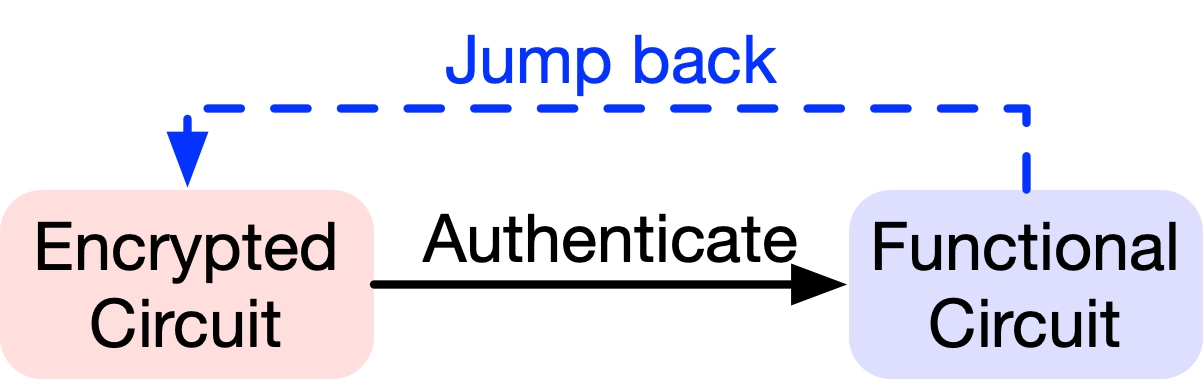}
}
}
\caption{Conventional (a) and proposed (b) authentication protocols for logic encryption.}
\label{fig:unlock_convension}
\end{figure}

As shown in Fig.~\ref{fig:unlock_convension}a, existing logic encryption techniques are mostly based on a single-authentication protocol, requiring users to be authenticated only once before using the correct circuit function. 
After authentication, the circuit remains functional unless it is powered off or reset. To attack the circuit, it is then sufficient to discover the correct key sequence that must be applied in the initial state. 

We adopt, instead, the authentication protocol in Fig.~\ref{fig:unlock_convension}b, 
where the functional circuit can ``jump'' back to the encrypted mode from the functional mode. 
Once the back-jumping occurs, another round of authentication is required to resume the normal operation. 
The back-jumping can be triggered multiple times and involve a different key sequence for each re-authentication step.  
The hardness of attacking this protocol stems from both the increased number of the key sequences to be produced and the uncertainty on the time at which each sequence should be applied. A new temporal dimension adds to the decryption procedure, which poses a significantly higher threshold to the attackers. 

\subsection{Overview of the Encryption Scheme}

\begin{figure}[t]
\centerline{\includegraphics[width=0.8\columnwidth]{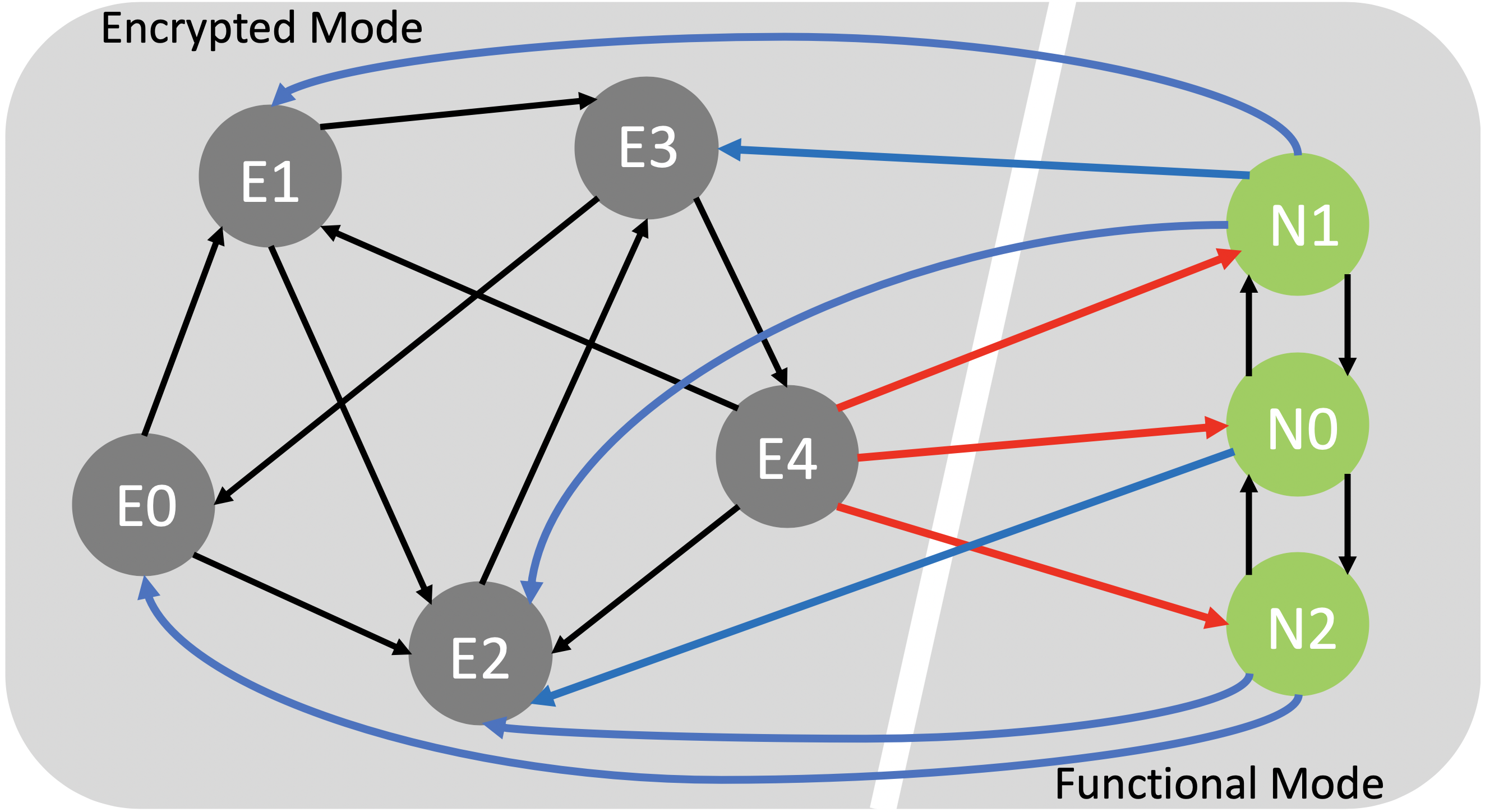}}
\caption{State transition diagram of SANSCrypt.}
\vspace{-15pt}
\label{fig:new_fsm}
\end{figure}

SANSCrypt is a sequential logic encryption scheme which supports random back-jumping, as represented in Fig.~\ref{fig:new_fsm}. 
When the circuit is powered or reset, the circuit falls into the reset state $E0$ of the encrypted mode. 
To transition to the initial (or reset) state $N0$ of the functional mode, the user must apply at startup the correct key sequence to the primary input ports. 

\begin{figure}[t]
\centerline{\includegraphics[width=0.9\columnwidth]{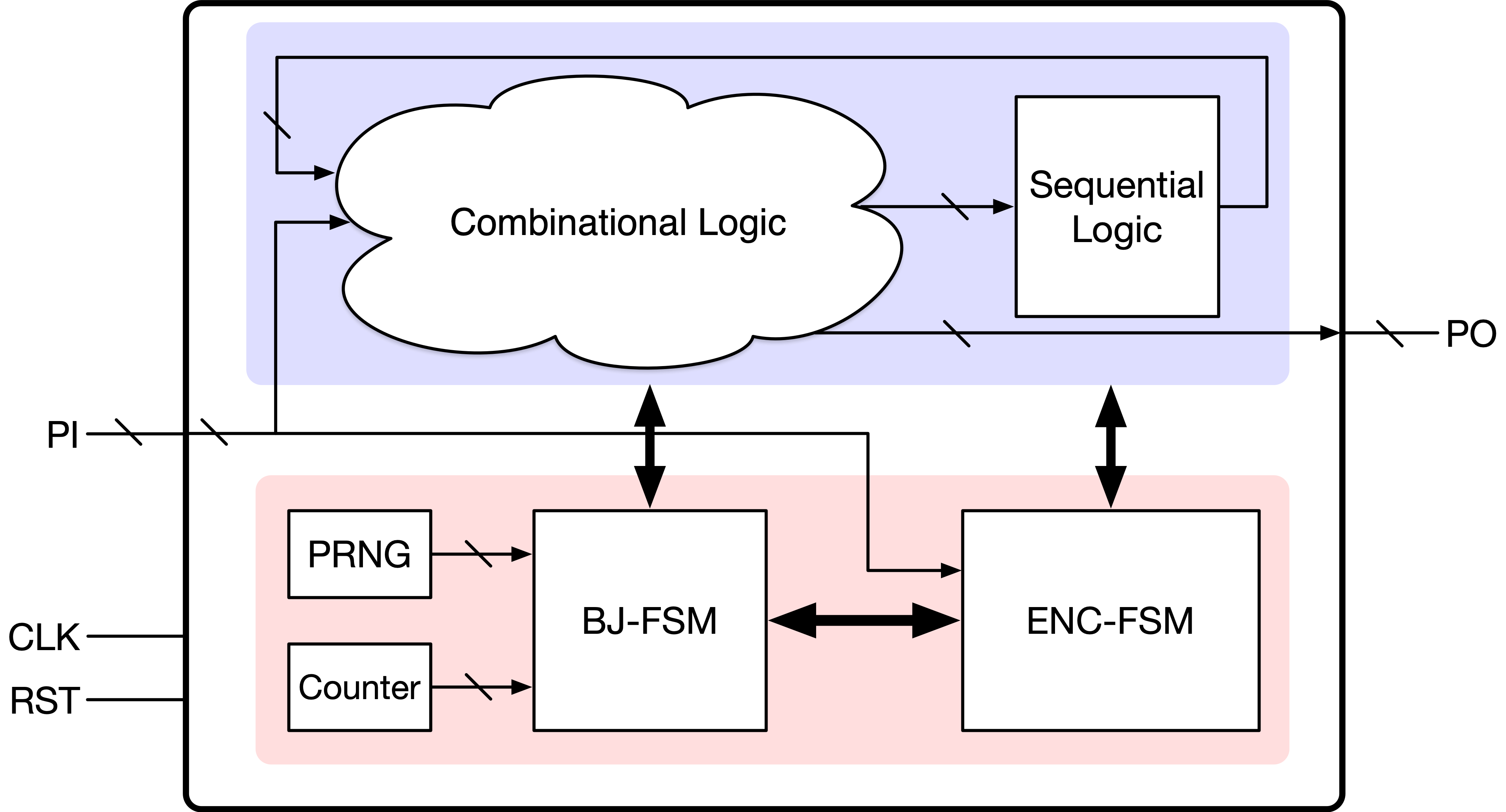}}
\caption{Schematic view of SANSCrypt.}
\label{fig:new_schematic}
\end{figure}

Once in the functional mode, the circuit can deliberately, but randomly, jump back, as denoted by the blue edges in Fig.~\ref{fig:new_fsm}, to a state $s_{bj}$ in the encrypted mode, called \textit{back-jumping state}, after a designated number of clock cycles $t_{bj}$, called \textit{back-jumping period}. 
The user needs to apply another key sequence to resume normal operations, as shown by the red arrows. 
Both the back-jumping state $s_{bj}$ and the back-jumping period $t_{bj}$ are determined by a pseudo-random number generator (PRNG) embedded in the circuit. 
Therefore, when and where the back-jumping operation happens is unpredictable unless the attacker is able to break the PRNG or find its seed. The schematic of SANSCrypt is shown in Fig.~\ref{fig:new_schematic} and consists of two additional blocks, a back-jumping module and an encryption finite state machine (ENC-FSM), besides the original circuit. 
We discuss each of these blocks in the following subsections. 

\subsection{Back-Jumping Module}

\begin{figure}[t]
\centerline{\includegraphics[width=0.9\columnwidth]{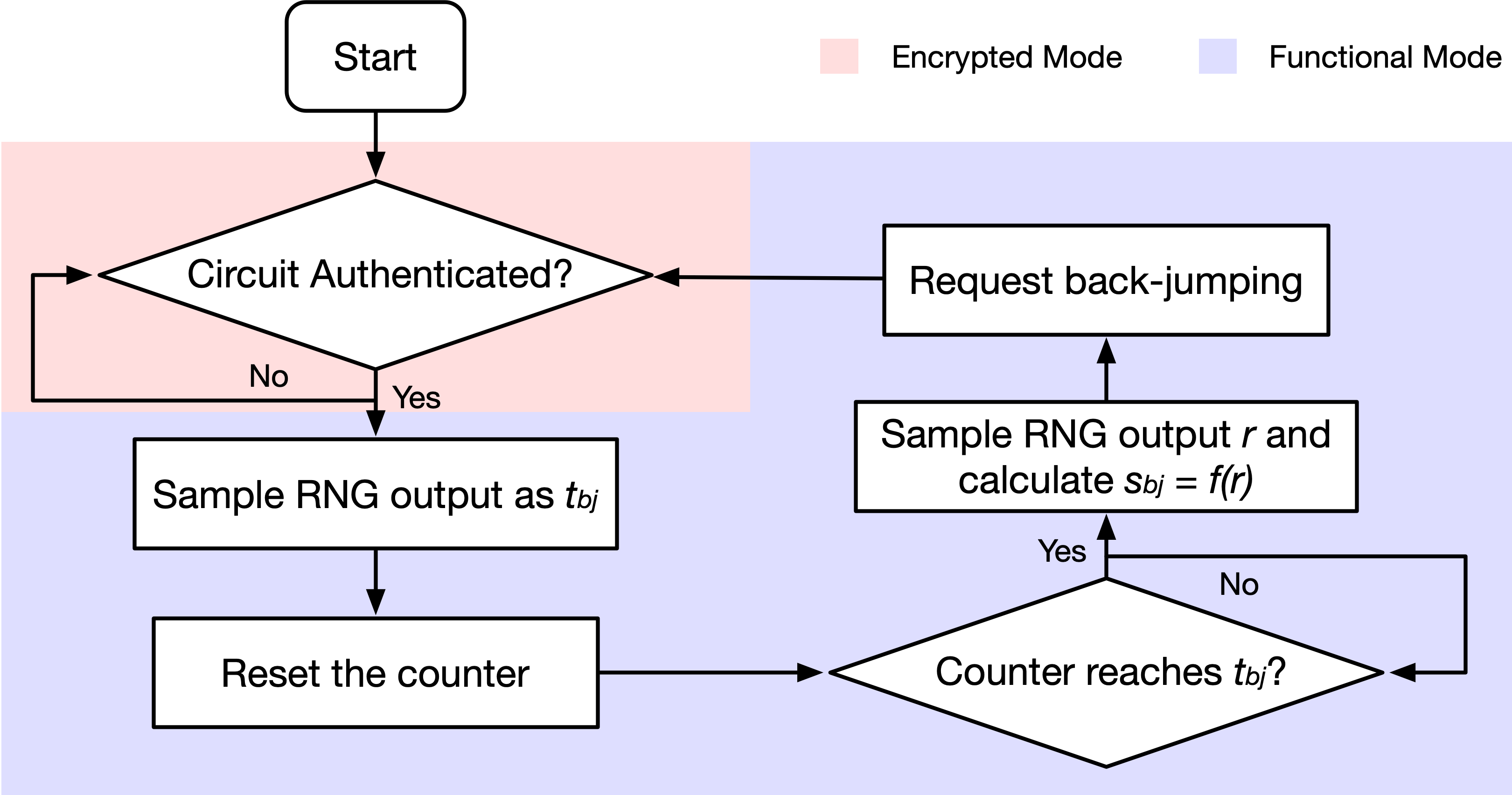}}
\caption{Flowchart of BJ-FSM. }
\vspace{-15pt}
\label{fig:bj_fsm}
\end{figure}

The back-jumping module consists of an $n$-bit \emph{PRNG}, an $n$-bit \emph{Counter}, and a \emph{Back-Jumping Finite State Machine} (BJ-FSM) which sends back-jumping commands to the rest of the circuit. As summarized in the flowchart in Fig.~\ref{fig:bj_fsm}, when the circuit is in the encrypted mode, BJ-FSM checks whether the  authentication has occurred. If this is the case, BJ-FSM stores the current PRNG output as the back-jumping period $t_{bj}$ and initializes the counter.

The counter increments its output at each clock cycle
until it reaches $t_{bj}$. This event triggers BJ-FSM to sample again the current PRNG output $r$, which is generally different from $t_{bj}$, and use it to determine the back-jumping state $s_{bj}= f(r)$. 
For example, if $s_{bj}$ is an $l$-bit binary number, BJ-FSM can arbitrarily select $l$ bits from $r$ and assign the value to $s_{bj}$. If the first $l$ bits of $r$ are selected, we have $f(r)=r[0:l-1]$.
At the same time, BJ-FSM sends a back-jumping request to the other blocks of the circuit and returns to its initial state, where it keeps checking the authentication status of the circuit. On receiving the back-jumping request, the circuit jumps back to state $s_{bj}$ in the encrypted mode and will stay there unless re-authentication is performed. 
Any PRNG architecture can be selected in this scheme, based on the design budget and the desired security level. For example, linear PRNGs, such as Linear Feedback Shift Registers (LFSRs), provide higher speed and lower area overhead but tend to be more vulnerable than cipher algorithm-based PRNGs, such as AES, which are, however, more expensive.

\subsection{Encryption Finite State Machine (ENC-FSM)}

\begin{figure}[t]
\centerline{\includegraphics[width=0.6\columnwidth]{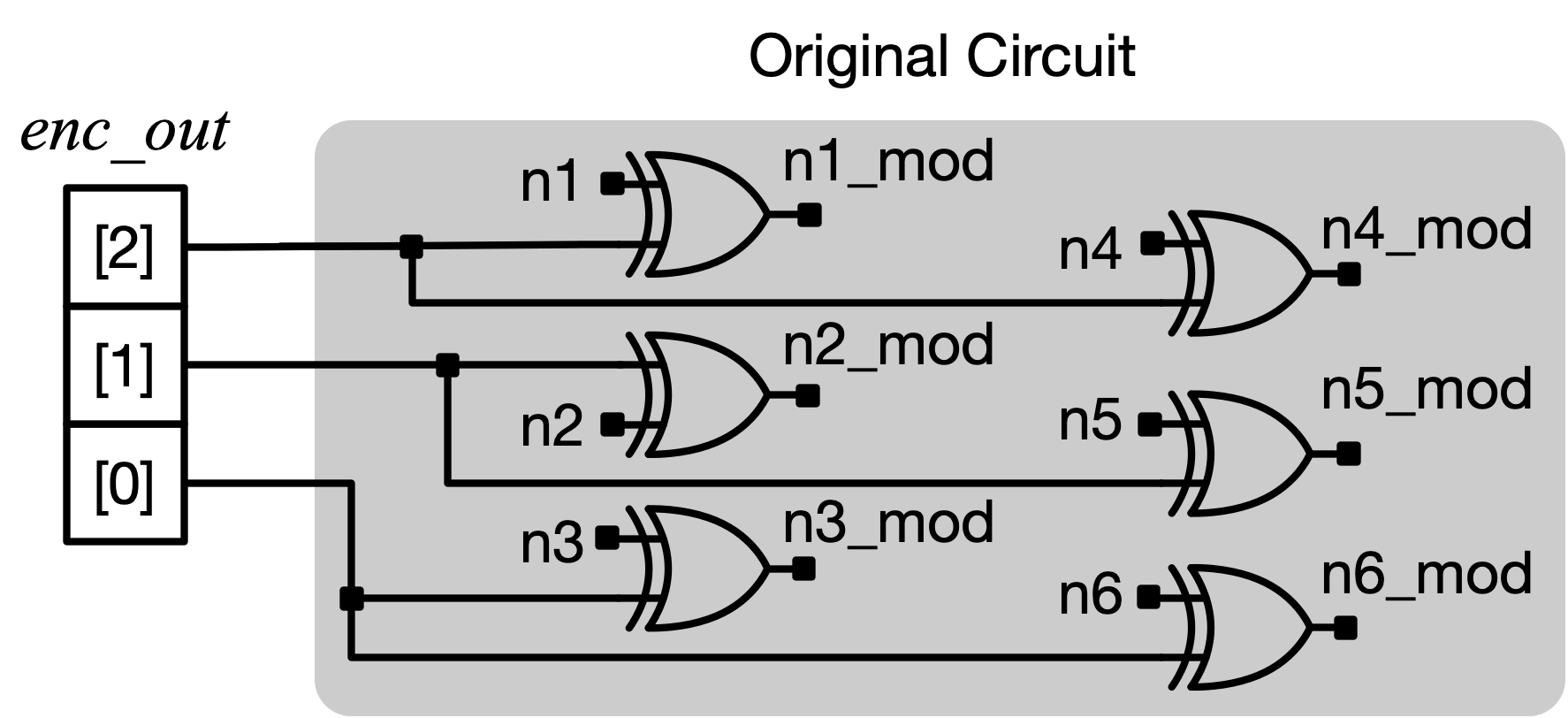}}
\caption{$enc\_out$ controls the original circuit via XOR gates.}
\label{fig:xor_corrupt}
\end{figure}

The Encryption Finite State Machine (ENC-FSM) determines whether the user's key sequence is correct and, if it is not correct, takes actions to hide the functionality of the original circuit. 
The input of the ENC-FSM can be provided via the primary input ports,
without the need to create extra input ports for authentication.  
The output $enc\_out$ of ENC-FSM, which is $n$ bit long, together with a set of nodes in the original circuit netlist, can be provided as an input to a set of XOR gates, to corrupt the circuit function as in combinational logic encryption~\cite{rajendran2013fault}. 
For example, in Fig.~\ref{fig:xor_corrupt},  a 3-bit array $enc\_out$ is connected to six nodes in the original circuit via XOR gates. In this paper, XOR gates are inserted at randomly selected nodes. However, any other combinational logic encryption technique is also applicable. As a design parameter, we denote by \textit{node coverage} the ratio between the number of inserted XOR gates and the total number of combinational logic gates in the circuit. 

Only one state of ENC-FSM, termed $auth$, is used in the functional mode. 
In state $auth$, all bits in $enc\_out$ are set to zero and the original circuit functionality is activated. 
In the other states, the value of $enc\_out$ changes based on the state, but at least one of its bits is set to one to guarantee that the final output is incorrect. 
A sample truth table for a 3-bit $enc\_out$ array is shown in Table~\ref{tab:enc_fsm_out}. Even if the circuit is in the encrypted mode, $enc\_out$ changes its value based on the state of the encryption FSM. Such an approach makes it difficult for signal analysis attacks, aiming to locate signals with low switching activity in the encrypted mode, to find 
$enc\_out$ and bypass ENC-FSM. 
After a valid authentication, the circuit resumes its normal operation. 
Additional registers are, therefore, required in the ENC-FSM to store the circuit state before back-jumping so that it can be resumed after authentication.

\begin{table}[t]
    \caption{Truth Table for a 3-Bit $enc\_out$ Array}
    \begin{center}
    \begin{tabular}{c|c|c|c|c|c|c}
    \textbf{State} & \textbf{E0} & \textbf{E1} & \textbf{E2} & \textbf{E3} & \textbf{E4} & \textbf{Auth} \\\hline
    $\mathbf{enc\_out[0]}$ & 0 & 1 & 1 & 1 & 1 & 0 \\\hline
    $\mathbf{enc\_out[1]}$ & 1 & 0 & 1 & 1 & 0 & 0 \\\hline
    $\mathbf{enc\_out[2]}$ & 1 & 1 & 1 & 0 & 0 & 0 \\ 
    \end{tabular}
    \label{tab:enc_fsm_out}
    \end{center}
\end{table}

\section{Performance Analysis}\label{sec:analysis}

We analyze SANSCrypt's resilience against existing attacks and estimate its overhead. 

\subsection{Brute-Force Attack}
Let us suppose that the number of primary input bits used as key inputs is $i$ and each re-authentication procedure requires $c$ clock cycles to apply the key sequence. 
If the attacker has no preference in selecting the key sequence, then she would have, on average, $(2^{i \cdot c}+1)/2\approx 2^{i \cdot c -1}$ attempts for each re-authentication procedure, which amounts to the same brute-force attack complexity of HARPOON. 
However, because the correct key sequence of each re-authentication procedure depends on the PRNG output, the number $N_{prng}$ of possible values of the PRNG output will also contribute to the attack effort. 
If each PRNG output corresponds to a unique key sequence which is independent from other key sequences, the average attack effort will be $N_{prng}\cdot 2^{i \cdot c -1}$. 
For a 10-bit PRNG, $i = 32$, and $c = 8$, the average attack effort will reach $5.93\times 10^{79}$. 

\subsection{Sequential SAT-Based Attack}

A SAT-based attack can be carried out on sequential encryption by unrolling the sequential portions of the circuit~\cite{shamsi2019kc2}. This attack can be remarkably successful especially when the correct key is the same at each time (clock cycle) and the key input ports are different from the primary input ports. Similarly to HARPOON, SANSCrypt is resilient to this SAT-based attack variant, since the correct keys are generally not the same at different clock cycles. 

\begin{figure}[t]
\centerline{\includegraphics[width=0.95\columnwidth]{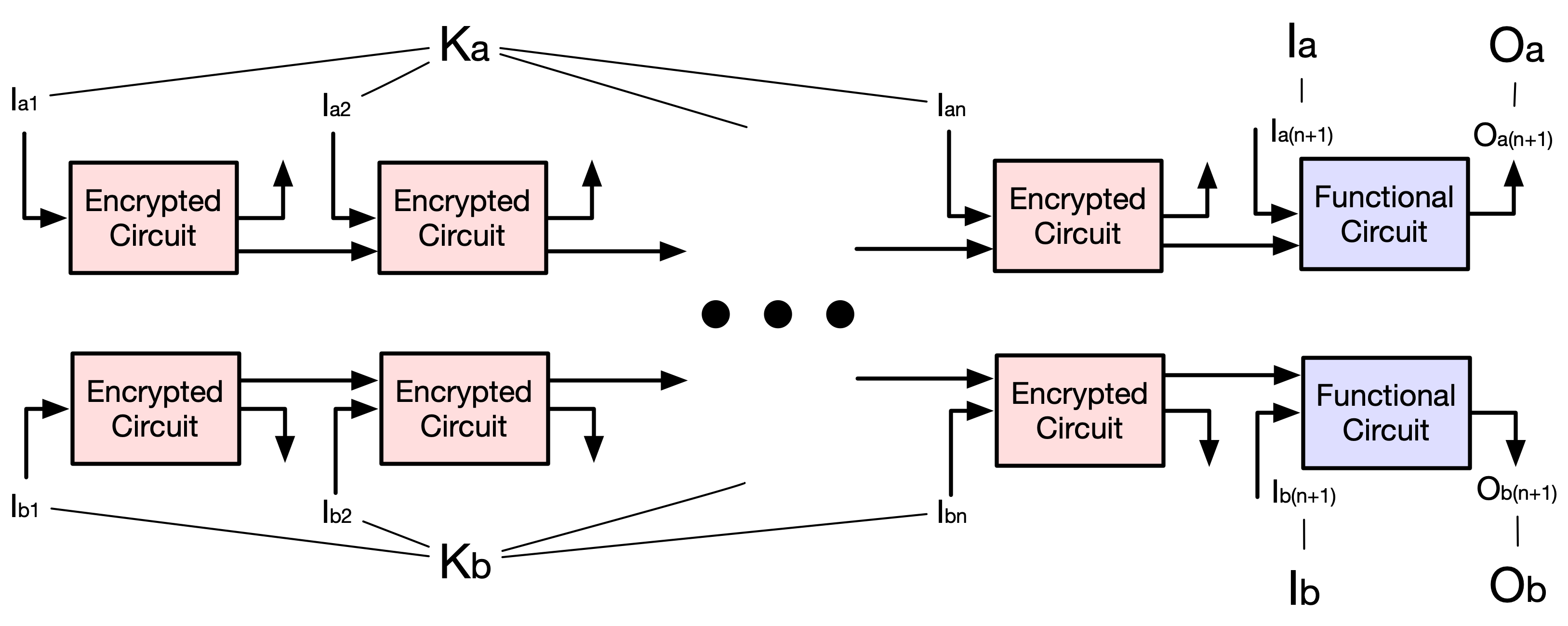}}
\caption{An unrolled encrypted circuit which requires $n$ clock cycles to find the key sequence.}
\label{fig:new_sat_attack}
\end{figure}

We therefore analyze the resilience of SANSCrypt via a modified version of the sequential SAT-based attack~\cite{meade2017revisit} that is appropriate for schemes such as HARPOON and SANSCrypt, as shown in Fig.~\ref{fig:new_sat_attack}.  
Let us first assume that the encryption scheme requires $n$ clock cycles after reset to enter the functional mode. Then, the attacker can start the attack by unrolling the circuit $(n+1)$ times. The first $n$ copies of the circuit receive the keys  at their primary input ports ($K_a$ and $K_b$), while the primary input and output ports of the $(n+1)^{th}$ circuit replica can be used to read the circuit input and output signals after $n$ cycles. If the SAT-based attack fails to find the correct key with $(n+1)$ circuit replicas, as in Fig.~\ref{fig:new_sat_attack}, the circuit will be unrolled one more time (see, e.g., \cite{shamsi2019kc2}). 

The attack above would be still ineffective on SANSCrypt, since it can retrieve the first key sequence but would fail to discover when  
the next back-jumping occurs and what would be the next key sequence. 
Even if the attacker knows when the next back-jumping occurs, the above SAT-based attack will fail due to the large number of circuit replicas needed to find all the key sequences, as empirically observed  in Section~\ref{sec:experiment}. 

\subsection{FSM Extraction and Structural Analysis}

As discussed in Section~\ref{sec:background}, a possible shortcoming of certain sequential encryption schemes is the clear boundary between the encrypted mode and the functional mode FSMs. 
As shown in Fig.~\ref{fig:new_schematic}, SANSCrypt addresses this issue by designing more than one transition between the two FSMs. 

An attacker may also try to locate and isolate the output of ENC-FSM by looking for low signal switching activities when the circuit is in the encrypted mode. 
SANSCrypt addresses this risk by expanding the output of ENC-FSM from one bit to an array. The value of each bit changes frequently based on the state of the encrypted mode FSM, which makes it difficult for attackers to find the output of ENC-FSM based on signal switching activities.

\subsection{Cycle Delay Analysis}

\begin{figure}[t]
\centerline{\includegraphics[width=1\columnwidth]{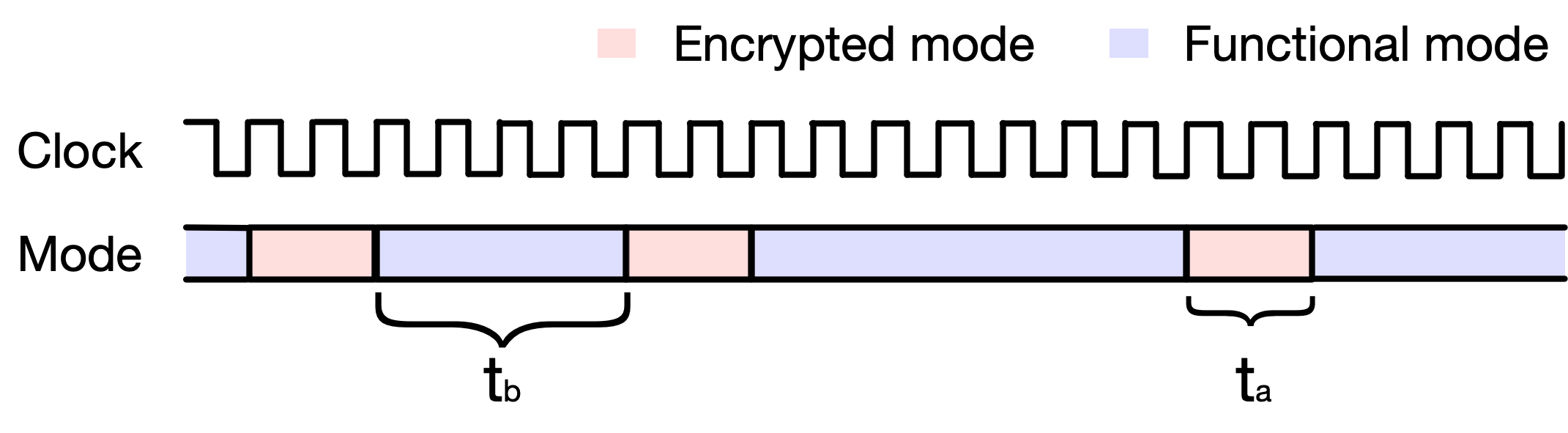}}
\vspace{-10pt}
\caption{Circuit mode switching for an authenticated user.}
\label{fig:waveform}
\end{figure}

Due to multiple back-jumping and authentication operations in SANSCrypt, additional clock cycles will be required in which no other operation can be executed. 
Suppose that authentication requires $t_{a}$ clock cycles 
and the circuit stays in the functional mode for $t_{b}$ clock cycles before the next back-jumping occurs, as shown in Fig.~\ref{fig:waveform}. 
The cycle delay overhead can be computed as the ratio $O_{cd} = {t_a}/{t_b}$.

\begin{figure}[t]
\centerline{\includegraphics[width=0.95\columnwidth]{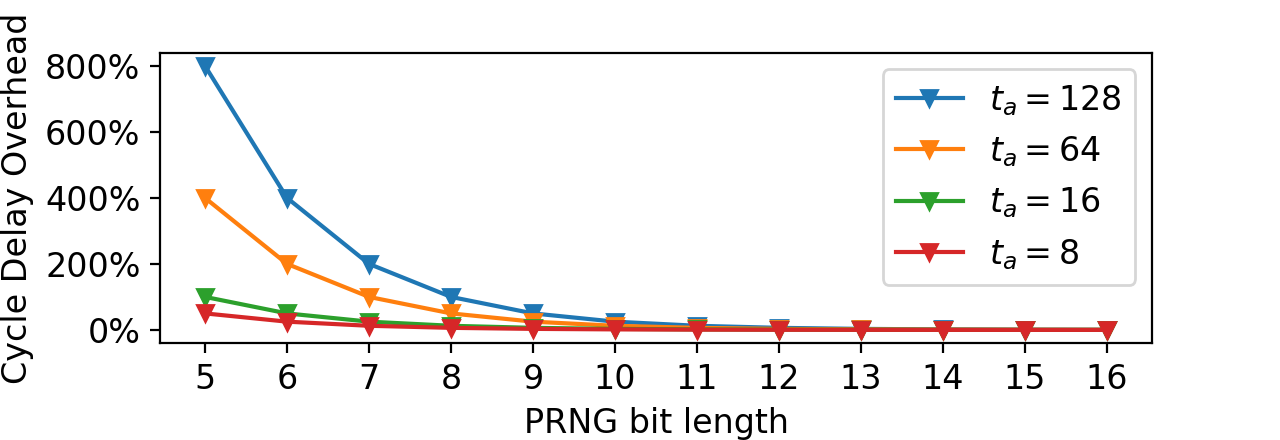}}
\caption{Average cycle delay as a function of PRNG bit length when the key sequence cycle length $t_{a}$ is 8, 16, 64, and 128.}
\label{fig:delay_curve}
\end{figure}

Specifically, for an $n$-bit PRNG, the average $t_{b}$ is equal to the average output value, i.e., $2^{n-1}$. 
To illustrate how the cycle delay overhead is influenced by this encryption, Fig.~\ref{fig:delay_curve} shows the relation between average cycle delay overhead and PRNG bit length. The clock cycles ($t_a$) required for (re-)authentication are set as 8, 16, 64, and 128.
When the PRNG bit length is small, the average cycle delay increases significantly with the increase of $t_{a}$. However, the cycle delay can be 
reduced by increasing the PRNG bit length. For example, the average cycle delay overhead becomes negligible for all the four cases when the PRNG bit length is 11 or larger. 
A key manager, available to the trusted user, will be in charge of automatically applying the key sequences from a tamper-proof memory at the right time, as computed from a hard-coded replica of the PRNG.

\begin{figure*}[t]
\centering
\subfigure[]{
\includegraphics[width=0.45\columnwidth]{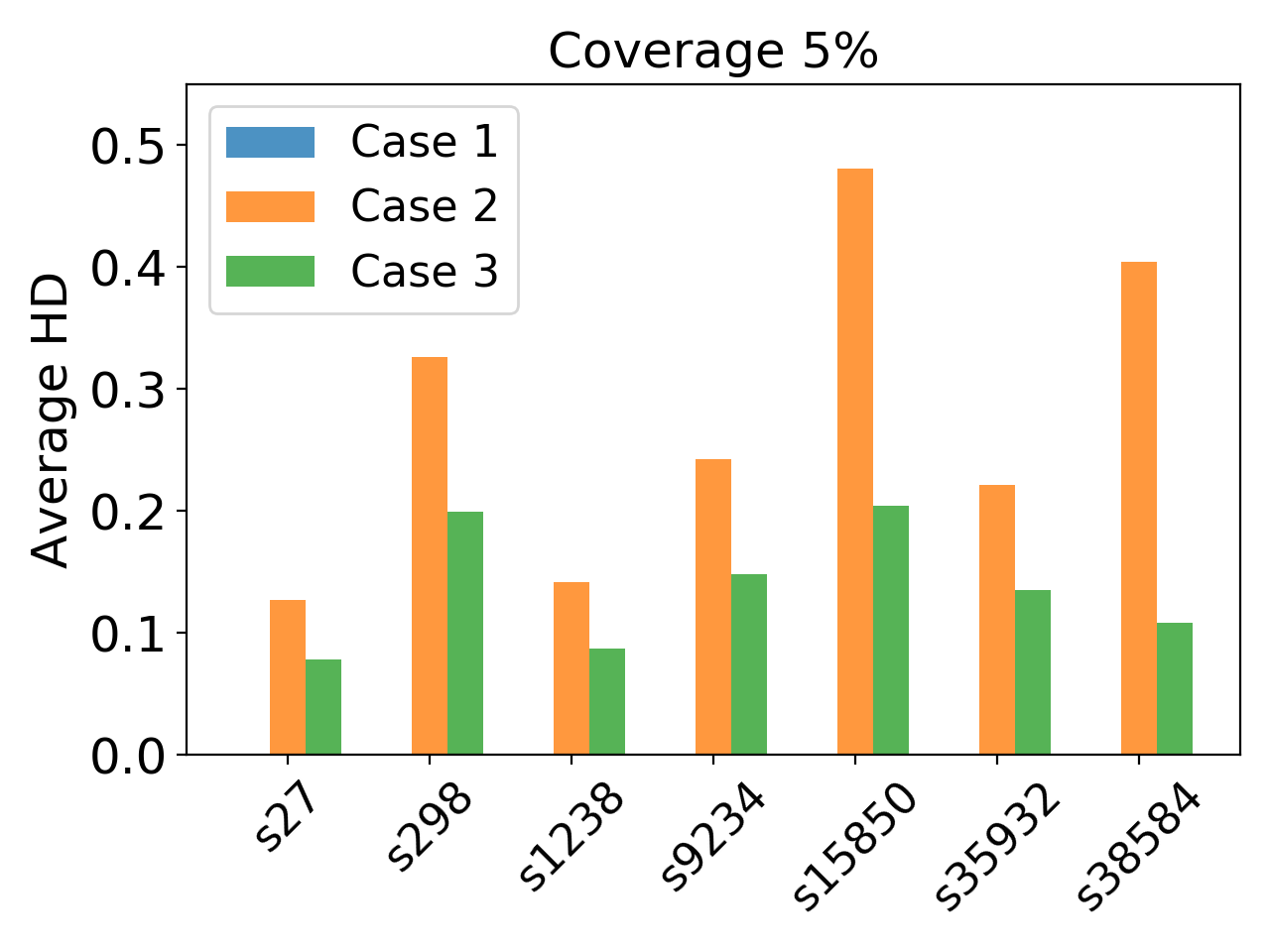}
}
\subfigure[]{
\includegraphics[width=0.45\columnwidth]{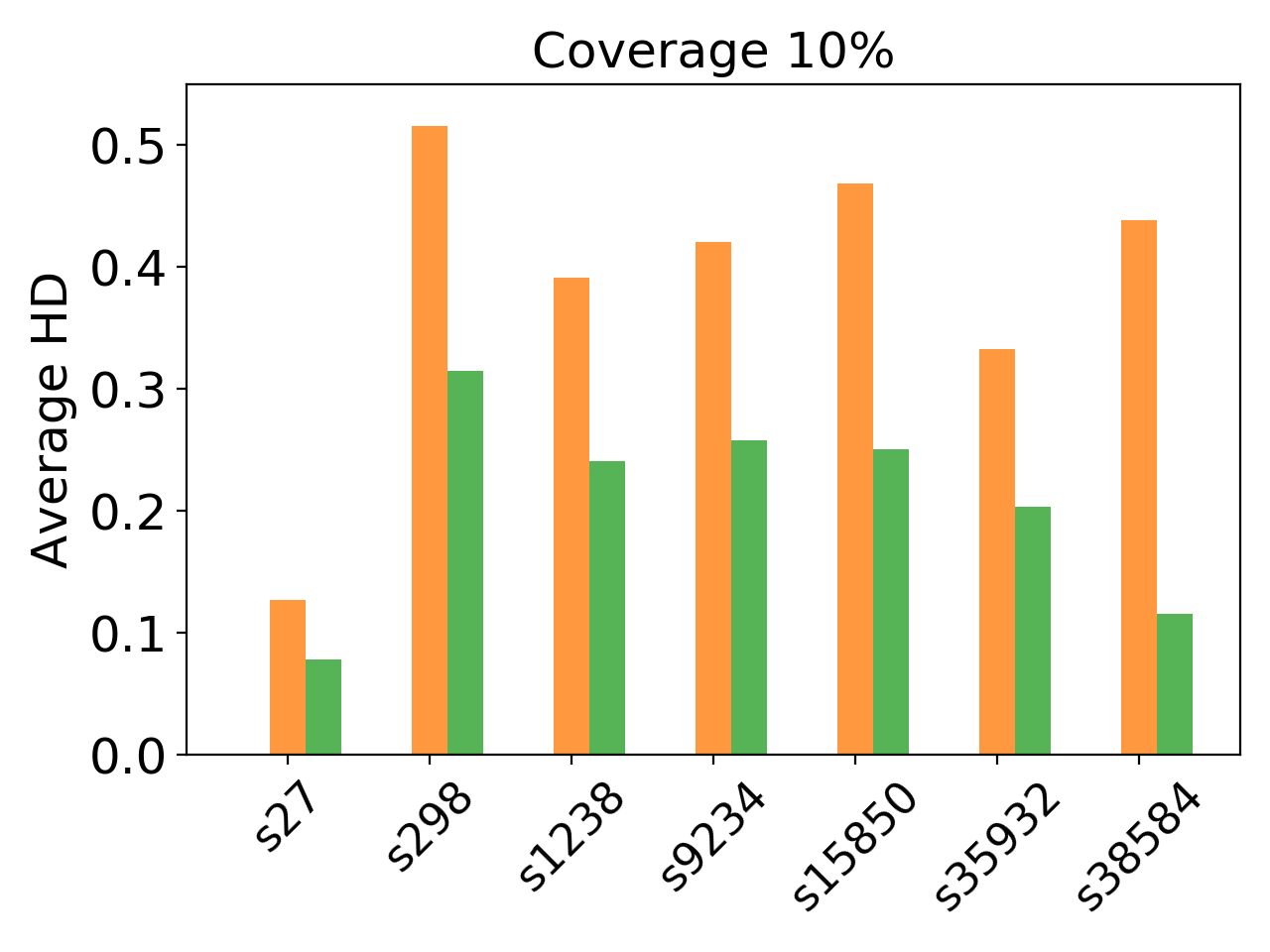}
}
\subfigure[]{
\includegraphics[width=0.45\columnwidth]{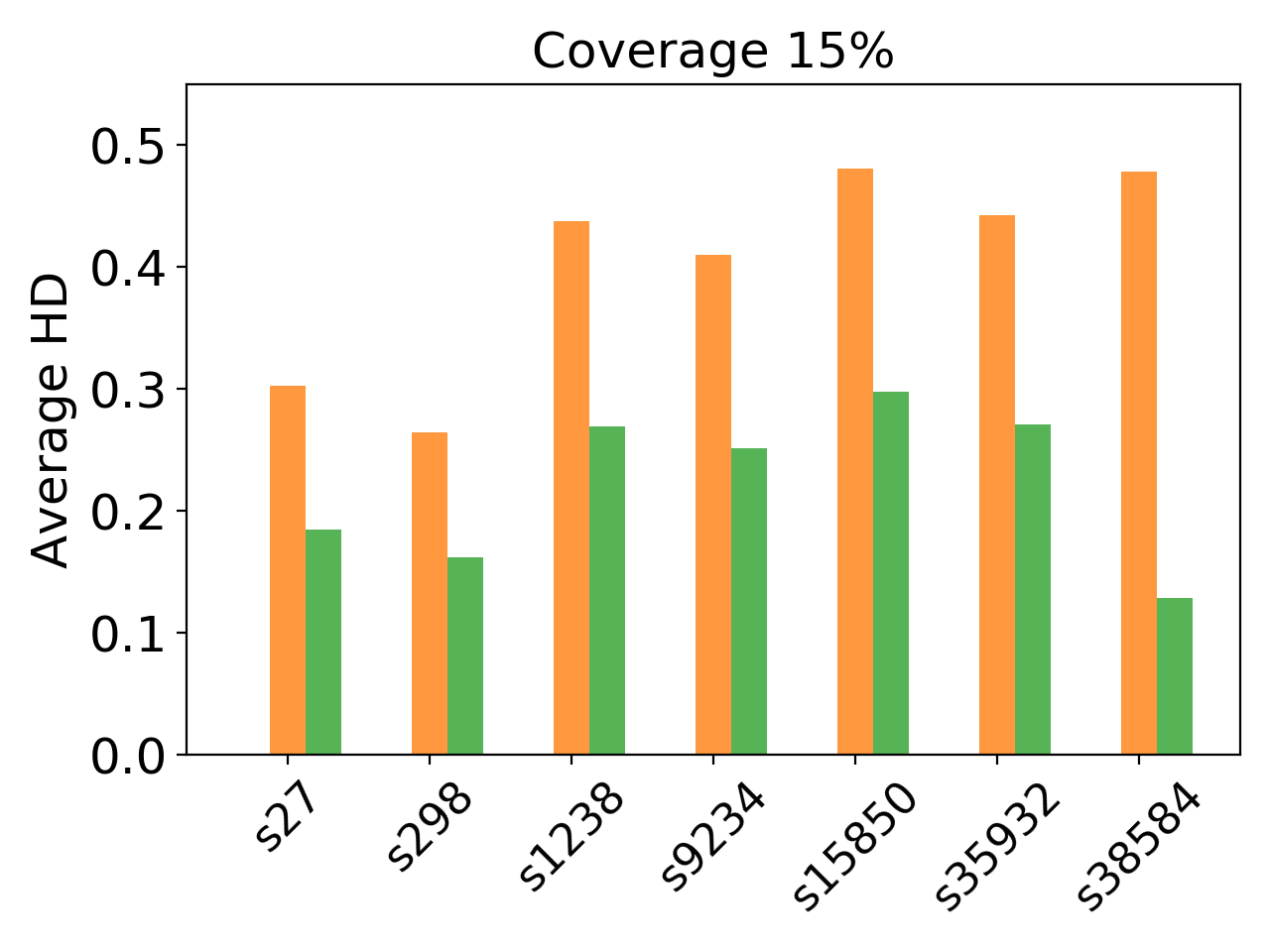}
}
\subfigure[]{
\includegraphics[width=0.45\columnwidth]{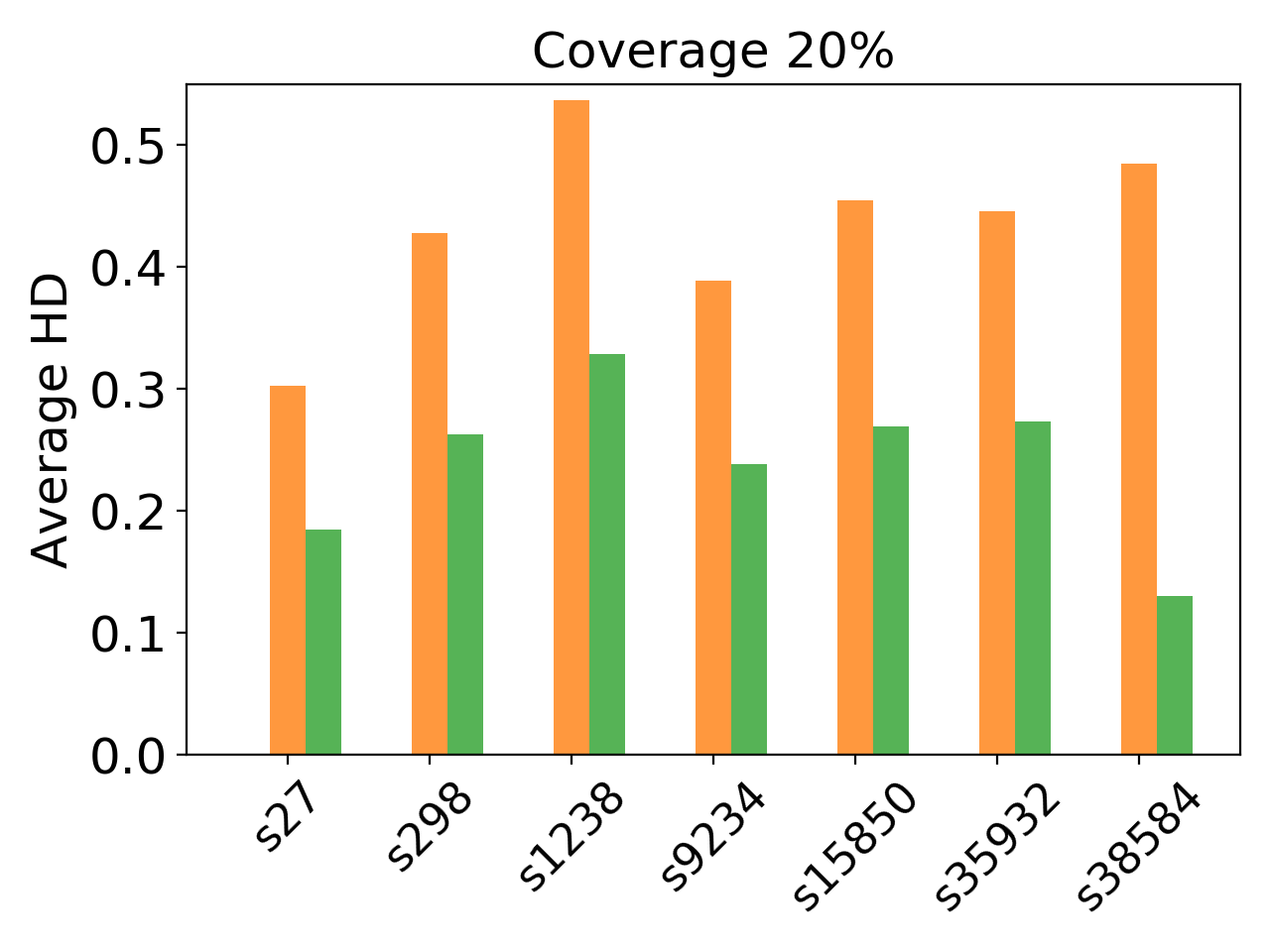}
}
\caption{The average HD for different node coverage: (a) $5\%$, (b) $10\%$, (c) $15\%$, and (d) $20\%$.}
\vspace{-10pt}
\label{fig:hd_comp}
\end{figure*}

\begin{table*}[t]
    \caption{Synthesis result of Area, Power, Delay}
    \begin{center}
    \resizebox{\textwidth}{!}{
    \begin{tabular}{|c|c|c|c|c|c|c|c|c|c|c|c|c|c|c|c|c|}
\hline
\textbf{Circuit}             & \multicolumn{4}{c|}{\textbf{s27}}   & \multicolumn{4}{c|}{\textbf{s298}}   & \multicolumn{4}{c|}{\textbf{s1238}}  & \multicolumn{4}{c|}{\textbf{s9234}}   \\ \hline
\textbf{Node Coverage} & 5\%  & 10\%  & 15\% & 20\% & 5\%   & 10\%  & 15\% & 20\% & 5\%   & 10\%  & 15\% & 20\% & 5\%   & 10\%   & 15\% & 20\% \\ \hline
\textbf{Area [\%]} & 1418.5  & 1418.5  & 1403.2  & 1403.2  & 413.0  & 427.3  & 425.2  & 453.8  & 144.8  & 165.7  & 176.0  & 189.2  & 114.6  & 131.7  & 144.5  & 160.1 \\ \hline
\textbf{Power [\%]} & 1627.7  & 1627.7  & 1627.5  & 1627.5  & 385.7  & 390.6  & 389.9  & 402.8  & 217.8  & 232.1  & 235.0  & 249.8  & 179.8  & 197.5  & 188.0  & 190.6 \\ \hline
\textbf{Delay [\%]} &0.0  & 0.0  & 1.4  & 1.4  & 0.0  & 0.0  & 0.0  & 0.5  & 0.0  & 0.0  & 0.0  & 5.8  & 0.0  & 0.0  & 0.9  & 3.6 \\ \hline
\textbf{Circuit}             & \multicolumn{4}{c|}{\textbf{s15850}} & \multicolumn{4}{c|}{\textbf{s35932}} & \multicolumn{4}{c|}{\textbf{s38584}} & \multicolumn{4}{c|}{\textbf{Average (s27 and s298 excluded)}} \\ \hline
\textbf{Node Coverage} & 5\%  & 10\%  & 15\% & 20\% & 5\%   & 10\%  & 15\% & 20\% & 5\%   & 10\%  & 15\% & 20\% & 5\%   & 10\%   & 15\% & 20\% \\ \hline
\textbf{Area [\%]} & 92.9  & 112.1  & 120.1  & 133.9  & 116.3  & 129.5  & 139.4  & 151.6  & 133.5  & 140.9  & 158.7  & 165.6 & 120.4  & 136.0  & 147.8  & 160.1 \\ \hline
\textbf{Power [\%]} & 127.4  & 142.3  & 153.2  & 163.0  & 98.4  & 101.9  & 101.2  & 103.0  & 123.9  & 128.8  & 142.0  & 140.3 & 149.5  & 160.5  & 163.9  & 169.4 \\ \hline
\textbf{Delay [\%]} & -0.3  & 0.0  & 0.1  & 0.6  & -0.4  & 0.0  & 4.3  & 5.3  & 0.6  & 2.0  & 0.4  & 4.9 & 0.0  & 0.4  & 1.1  & 4.0 \\ \hline
    \end{tabular}}
    \label{tab:overhead}
    \end{center}
\end{table*}

\section{Simulation Results}
\label{sec:experiment}

We first evaluate the effectiveness of SANSCrypt on seven ISCAS'89 sequential benchmark circuits with different sizes, as summarized in Table~\ref{tab:benchmark}. All the experiments are executed on a Linux server with 48 $2.1$-GHz processor cores and 500-GB memory. We implement our technique on the selected circuits with different configurations and use a 45-nm NangateOpenCellLibrary~\cite{nangate} to synthesize the encrypted netlists for area optimization under a critical-path delay constraint that targets the same performance as in the non-encrypted versions. 
\begin{table}[t]
    \caption{Overview of the Selected Benchmark Circuits}
    \begin{center}
    \resizebox{\columnwidth}{!}{
    \begin{tabular}{|c|c|c|c|c|c|c|c|}
    \hline
    \textbf{Circuit} & \textbf{s27} & \textbf{s298} & \textbf{s1238} & \textbf{s9234} & \textbf{s15850} & \textbf{s35932} & \textbf{s38584} \\\hline
    \textbf{Input} & 4 & 3 & 14 & 36 & 77 & 35 & 38 \\\hline
    \textbf{Output} & 1 & 6 & 14 & 39 & 150 & 320 & 304 \\\hline
    \textbf{DFF} & 3 & 14 & 18 & 211 & 534 & 1728 & 1426 \\\hline
    \textbf{Gate} & 10 & 119 & 508 & 5597 & 9772 & 16065 & 19253 \\\hline
    \end{tabular}}
    \label{tab:benchmark}
    \end{center}
\end{table}
For the purpose of illustration, we realize the PRNG using Linear Feedback Shift Registers (LFSR) with different sizes, ranging from 5 to 15 bits. An LFSR provides an area-efficient implementation and has often been used in other logic encryption schemes in the literature~\cite{rahmandynamically,xiao2015efficient}. 
We choose a random 8-cycle-long key sequence as the correct key, and select $5\%$, $10\%$, $15\%$, and $20\%$ as node coverage levels.  
Finally, we use the Hamming distance (HD) between the correct  and the corrupted output values as a metric for the output corruptibility. If the HD is 0.5, 
the effort spent to identify the incorrect bits is maximum. 

We run functional simulations on all the encrypted circuits with the correct key sequences (case 1) and without the correct sequences (case 2), by applying 1000 random input vectors. 
We then compare the circuit output with the golden output from the original netlist and calculate the HD between the two. Moreover, we demonstrate the additional robustness of SANSCrypt by simulating a scenario (case 3) in which the attacker assumes that the encryption is based on a single-authentication protocol and provides only the first correct key sequence upon reset. 
Fig.~\ref{fig:hd_comp}a-d show the average HD in these three cases.
For all the circuits, the average HD is zero only in case 1, when all the correct key sequences are applied at the right clock cycles. Otherwise, in case 2 (orange) and case 3 (green), we observe a significant increase in the average HD. 
The average HD in case 3 is always smaller than that of case 2 because, in case 3, the correct functionality is recovered for a short period of time, after which the circuit jumps back to the encrypted mode. The longer the overall runtime, the smaller will be the impact of the transparency window in which the circuit exhibits the correct functionality. 

We then apply the sequential SAT-based attack in Section~\ref{sec:analysis} to circuit s1238 with 5-bit LFSR and 20\% node coverage, under a stronger attack model, in which the attacker knows when to apply the correct key sequences. Table~\ref{tab:sat_attack} shows the runtime to find the first set of 7 key sequences. 
\begin{table}[t]
    \caption{SAT-based attack runtime for finding the first 7 key sequences}
    \begin{center}
    \resizebox{\columnwidth}{!}{
    \begin{tabular}{|c|c|c|c|c|c|c|c|}
    \hline
    \textbf{Key Seq. Index} & 1 (HARPOON) & 2 & 3 & 4 & 5 & 6 & 7 \\\hline
    \textbf{Runtime [s]} & 4 & 123 & 229 & 1941 & 1301 & 2202 & 25571 \\\hline
    \end{tabular}}
    \label{tab:sat_attack}
    \end{center}
\end{table}
The runtime remains exponential in the number of key sequences, which makes sequential SAT-based attacks impractical for large designs.

Finally, Table~\ref{tab:overhead} reports the synthesized area, power, and delay overhead due to the implementation of our technique. 
In more than $70\%$ of the circuits the delay overhead is less than $1\%$, and exceeds the required clock cycle by at most $5.8\%$. Except for \emph{s27} and \emph{s298}, characterized by a small gate count, all the other circuits show average area and power overhead of $141.1\%$ and $160.8\%$, respectively, which is expected due to the additional number of registers required in ENC-FSM to guarantee that the correct state is entered upon re-authentication. However, because critical modules in large SoCs may only account for a small portion of the area, this overhead becomes affordable under partial obfuscation. 
For example, we encrypted a portion of state registers in \emph{s38584}, the largest ISCAS'89 benchmark, using SANSCrypt. We then randomly inserted additional XOR gates to achieve the same HD as in the case of full encryption.
Table~\ref{tab:partial_enc} reports the overhead results after synthesis, when the ratio between the encrypted state registers and the total number of state registers decreases from $100\%$ to $1\%$. 
\begin{table}[t]
    \caption{ADP Overhead Results for Partial Encryption}
    \begin{center}
    \resizebox{\columnwidth}{!}{
    \begin{tabular}{|c|c|c|c|c|c|c|c|}
    \hline
    \textbf{Encrypted registers/Total registers} & 100\% & 50\% & 25\% & 10\% & 5\% & 2.5\% & 1\% \\\hline
    \textbf{Area [\%]} & 133.5 & 71.6 & 49.1 & 33.4 & 27.8 & 23.5 & 22.4 \\\hline
    \textbf{Power [\%]} & 123.9 & 40.2 & 9.6 & -12.8 & -20.5 & -22.1 & -25.0 \\\hline
    \textbf{Delay [\%]} & 0.6 & 1.8 & 2.1 & 4.2 & 5.4 & 3.9 & 4.6 \\\hline
    \end{tabular}}
    \label{tab:partial_enc}
    \end{center}
\end{table}
Encrypting $10\%$ of the registers will only cost $33.4\%$ of the area while incurring negative power overhead and $4.2\%$ delay overhead.

\section{Conclusion}\label{sec:conclusion}

We proposed SANSCrypt, a robust sequential logic encryption technique relying on a  sporadic authentication protocol, in which re-authentications are carried out at pseudo-randomly selected time slots to significantly increase the attack effort. 
Future work includes optimizing the implementation to further reduce the overhead and hide any structural traces that may expose the correct key sequence. Further, we plan to investigate key manager architectures to guarantee reliable timing and operation in real-time applications. 

\section*{Acknowledgment}
This work was partially sponsored by the Air Force Research Laboratory (AFRL) and the Defense Advanced Research Projects Agency (DARPA) under agreement number FA8560-18-1-7817.

\bibliographystyle{ieeetr}
{\small
\bibliography{reference_list}}
\end{document}